\documentclass[12pt,a4paper]{article}
\usepackage{amsfonts}
\usepackage{amssymb}
\usepackage{amsmath}
\usepackage{latexsym}
\begin{document}

\begin{center}
{\Large \bf Doubling of background solution\\ \vspace{2mm} in 5D
stabilized brane world model} \\

\vspace{4mm}

M.N.~Smolyakov\\
\vspace{0.5cm} Skobeltsyn Institute of Nuclear Physics, Moscow
State University,
\\ 119991, Moscow, Russia\\
\end{center}

\begin{abstract}
We discuss a model providing two different stationary background
solutions with flat and $dS_{4}$ metric on the branes under the
same values of the fundamental parameters. It is shown that only
an additional fine-tuning of the brane scalar field potentials can
provide a separation between two background solutions.
\end{abstract}

Stabilized brane world models have been widely discussed in the
last years. The most consistent model with flat metric on the
branes was proposed in paper \cite{DeWolfe}, where exact solutions
to equations of motion for the background metric and the scalar
field were found. The size of the extra dimension is defined by
the boundary conditions for the scalar field on the branes.

Most brane world models assume the metric on the branes to be the
flat Minkowski metric. At the same time it is evident that more
realistic models should account for a cosmological evolution on
the branes. This problem is widely discussed in scientific
literature, see, for example \cite{DeWolfe,Barger:2000wj}, reviews
\cite{Brax:2003fv,Brax:2004xh} and references therein. Quite an
interesting class of the brane world models is the one describing
background solutions with $dS_{4}$ metric on the branes. Different
models with the scalar field living in the bulk and $dS_{4}$
metric on the branes were discussed in
\cite{DeWolfe,Cline:2000ky,Kanti:2002ta}.

In this paper we discuss a model which provides different exact
solutions (with flat Minkowski or $dS_{4}$ background metric on
the branes) for different values of the fine-tuned brane tensions
in both cases (the same problem was discussed in
\cite{Cline:2000ky}, where an exact solution with zero Hubble
parameter on the branes and an approximate solution with nonzero
constant Hubble parameter on the branes were obtained for the same
bulk scalar field potential). We also show that under an
appropriate value of the parameter of the bulk scalar field
potential the equations of motion lead to these different exact
solutions even for the same values of the brane tensions in both
cases.

To this end let us consider gravity in a five-dimensional
space-time $E=M_4\times S^1/Z_2$, interacting with two branes and
with the scalar field $\phi$. Let us denote coordinates in $E$ by
$\{x^M\}=\{t,x^i,y\}$, $ M=0,1,2,3,4$, where $x^{0}\equiv t$;
$\{x^i\},\: i=1,2,3$ are three-dimensional spatial coordinates and
the coordinate $y\equiv x^4$, $-L\le y\le L$, corresponds to the
extra dimension. The extra dimension forms the orbifold $S^1/Z_2$,
which is a circle of diameter $2L/\pi$ with the points $y$ and
$-y$ identified. Correspondingly, the metric $g_{MN}$ and the
scalar field $\phi$ satisfy the orbifold symmetry conditions
\begin{eqnarray}\label{eq}
g_{\mu\nu}(x,-y)=g_{\mu\nu}(x,y)\,, \qquad
g_{\mu4}(x,-y)=-g_{\mu4}(x,y)\,, \\ \nonumber
g_{44}(x,-y)=g_{44}(x,y)\,, \qquad \phi(x,-y) =\phi(x,y)\,,
\end{eqnarray}
$\mu=0,1,2,3$. The branes are located at the fixed points of the
orbifold $y=0$ and $y=L$.

The action of the model has the form
\begin{eqnarray}\label{s1}
S=M^{3}\int R \sqrt{-g}d^5{x} -\int\left(\frac{1}{2}\partial_M
\phi
\partial^M \phi +  V(\phi)\right)\sqrt{-g} d^5{x}-\\ \nonumber
-\int\left(\lambda_1(\phi)\delta(y)
+\lambda_2(\phi)\delta(y-L)\right)\sqrt{-\tilde g}d^5{x}\,,
\end{eqnarray}
where $M$ is the five-dimensional Planck mass,
$\lambda_{1,2}(\phi)$ are the scalar field potentials on the
branes and $\tilde g_{\mu\nu}$ is the induced metric on the
branes.

We consider the standard form of the background metric, which is
often used in brane world models (see, for example,
\cite{DeWolfe})
\begin{equation}\label{backgmetric}
ds^2=\gamma_{MN}dx^{M}dx^{N}=e^{2A(y)}\left(-dt^{2}+a^{2}(t)\eta_{ij}dx^{i}dx^{j}\right)+dy^2\,,
\end{equation}
where $\eta_{ij}=diag(1,1,1)$ and
\begin{equation}\label{backgscalar}
\phi(x,y)=\phi(y)\,.
\end{equation}

Below we will consider a maximally-symmetric metric on the branes
with
\begin{equation}\label{backgHubble}
a(t)=e^{Ht}\,,
\end{equation}
where $H$ is the four-dimensional Hubble parameter. In this case
the equations of motion, following from (\ref{s1}), take the form
\cite{DeWolfe}
\begin{eqnarray}\label{1}
\phi''+4A'\phi'&=&\frac{dV}{d\phi}+\frac{d\lambda_1}{d\phi}\delta(y)+\frac{d\lambda_2}{d\phi}\delta(y-L)\,,\\
\label{2}
M^{3}\left({A'}^{2}-H^{2}e^{-2A}\right)&=&\frac{{\phi'}^{2}}{24}-\frac{V}{12}\,,\\
\label{3}
3M^{3}\left(A''+H^{2}e^{-2A}\right)&=&-\frac{{\phi'}^{2}}{2}-\frac{1}{2}\lambda_1\delta(y)-\frac{1}{2}\lambda_2\delta(y-L)\,.
\end{eqnarray}
We note that the equations for $00$-component of the Einstein
equations and $ij$-component of the Einstein equations have an
equal form and lead to (\ref{3}).

Equations (\ref{1}), (\ref{2}), (\ref{3}) can be rewritten as
equations on the interval $(0,L)$
\begin{eqnarray}\label{10}
\phi''+4A'\phi'&=&\frac{dV}{d\phi}\,,\\
\label{20}
M^{3}\left({A'}^{2}-H^{2}e^{-2A}\right)&=&\frac{{\phi'}^{2}}{24}-\frac{V}{12}\,,\\
\label{30}
3M^{3}\left(A''+H^{2}e^{-2A}\right)&=&-\frac{{\phi'}^{2}}{2}
\end{eqnarray}
and standard boundary conditions on the branes
\begin{eqnarray}\label{b1}
\phi'|_{y=+0}=\frac{1}{2}\frac{d\lambda_1}{d\phi}\,, \qquad
\phi'|_{y=L-0}=-\frac{1}{2}\frac{d\lambda_2}{d\phi}\,,\\
\label{b3} A'|_{y=+0}=-\frac{1}{12M^{3}}\lambda_1\,, \qquad
A'|_{y=L-0}=\frac{1}{12M^{3}}\lambda_2\,.
\end{eqnarray}
It is not difficult to show that only two equations of (\ref{10}),
(\ref{20}), (\ref{30}) are independent; one can show it by
differentiating (\ref{20}) and substituting the result into
(\ref{30}).

We consider the standard exponential bulk potential
\begin{equation}\label{potential}
V=-\beta^{2}e^{-\gamma\phi}
\end{equation}
and the following brane potentials
\begin{eqnarray}\label{potent1}
\lambda_{1}(\phi)=\epsilon_{1}+F_{1}(x)\cdot
\left(\phi-\phi_{1}\right),\\ \label{potent2}
\lambda_{2}(\phi)=\epsilon_{2}+F_{2}(x)\cdot
\left(\phi-\phi_{2}\right),
\end{eqnarray}
where $F_{1,2}(x)$ are auxiliary scalar fields, $\phi_{1,2}$ and
$\epsilon_{1,2}$ are constants. Note that the fields $F_{1,2}(x)$
have no kinetic terms. One can recall that supersymmetry is based
on the use of such auxiliary fields, which are necessary for
reaching the closure of the supersymmetry algebra \cite{Ramond}. A
simple example with the fields of such type in classical field
theory can be also found in \cite{Ramond}. The equations of
motions for the fields $F_{1,2}(x)$ (which can be obtained by
means of the standard variation procedure with respect to the
fields $F_{1,2}(x)$) give
\begin{eqnarray}\label{brane-phi1}
\phi|_{y=0}=\phi_{1}\,,\\ \label{brane-phi2}
\phi|_{y=L}=\phi_{2}\,.
\end{eqnarray}
In fact the fields $F_{1,2}(x)$ play the role of Lagrange
multipliers. We will see below that conditions (\ref{brane-phi1})
and (\ref{brane-phi2}) fix the size of the extra dimension. Such
method of stabilization (with the help of auxiliary fields) seems
to be quite simple and reduces the number of parameters to be
fine-tuned. The physical consequences are equal to those of the
stiff brane potentials used in \cite{Cline:2000ky}.

Now let us consider two possible cases.
\begin{enumerate}
\item $H=0$.\\
Solution to equations of motion (\ref{10}), (\ref{20}) and
(\ref{30}) in the interval $(0,L)$ has the form
\begin{eqnarray}
A&=&\frac{2}{3\gamma^{2}M^{3}}\ln(ky+C_{1})+C\,,\\ \label{ph1}
\phi&=&\frac{2}{\gamma}\ln(ky+C_{1})\,,
\end{eqnarray}
where $C$, $C_{1}$ are constants and
\begin{equation}
k^{2}=\frac{3\beta^{2}\gamma^{4}M^{3}}{16-6\gamma^{2}M^{3}}\,.
\end{equation}
We also suppose that $\gamma<2\sqrt{\frac{2}{3M^{3}}}$.

Equation (\ref{brane-phi1}) defines the constant $C_{1}$
$$C_{1}=e^{\frac{\gamma\phi_{1}}{2}}\,,$$
whereas the size of the extra dimension is defined by equation
(\ref{brane-phi2})
$$L=\frac{e^{\frac{\gamma\phi_{2}}{2}}-e^{\frac{\gamma\phi_{1}}{2}}}{k}\,.$$
The constant $C$ should be defined by the requirement to have
Galilean coordinates on the brane. It means that the values of $C$
are different for different branes: $C=-\frac{\phi_{1}}{3\gamma
M^{3}}$ for the brane at $y=0$ ($A(0)=0$) and
$C=-\frac{\phi_{2}}{3\gamma M^{3}}$ for the brane at $y=L$
($A(L)=0$), see detailed discussion about Galilean coordinates on
the branes in \cite{Rubakov,Boos,Kubyshin}. In other words, its
value is defined by the physical four-dimensional scale we suppose
to use.

We see that all the parameters of the background solution appear
to be defined. Boundary conditions (\ref{b1}) result in
\begin{eqnarray}\label{F1}
F_{1}&=&\frac{4k}{\gamma}\,e^{-\frac{\gamma\phi_{1}}{2}}\,,\\
\label{F2}
F_{2}&=&-\frac{4k}{\gamma}\,e^{-\frac{\gamma\phi_{2}}{2}}\,.
\end{eqnarray}
Boundary conditions (\ref{b3}) suggest that the brane tensions
$\epsilon_{1,2}$ should be fine-tuned
\begin{eqnarray}\label{E1}
\epsilon_{1}&=&-\frac{8k}{\gamma^{2}}e^{-\frac{\gamma\phi_{1}}{2}}=-12M^{3}\frac{\beta\gamma}{\sqrt{6}}
\frac{e^{-\frac{\gamma\phi_{1}}{2}}}{\sqrt{2M^{3}\gamma^{2}-\frac{3}{4}\gamma^{4}M^{6}}}\,,\\
\label{E2}
\epsilon_{2}&=&\frac{8k}{\gamma^{2}}e^{-\frac{\gamma\phi_{2}}{2}}=12M^{3}\frac{\beta\gamma}{\sqrt{6}}
\frac{e^{-\frac{\gamma\phi_{2}}{2}}}{\sqrt{2M^{3}\gamma^{2}-\frac{3}{4}\gamma^{4}M^{6}}}\,.
\end{eqnarray}
Such a fine-tuning is inherent to almost all five-dimensional
brane world models with compact extra dimension.

\item $H\ne 0$.\\
In this case the solution to equations of motion (\ref{10}),
(\ref{20}) and (\ref{30}) in the interval $(0,L)$ has the form
\begin{eqnarray}
A&=&\ln(ky+C_{1})+C\,,\\
\label{ph2} \phi&=&\frac{2}{\gamma}\ln(ky+C_{1})\,,
\end{eqnarray}
with
\begin{equation}
k^{2}=\frac{\beta^{2}\gamma^{2}}{6}
\end{equation}
and
\begin{equation}
H^{2}e^{-2C}=\frac{k^{2}}{3\gamma^{2}M^{3}}\left(3\gamma^{2}M^{3}-2\right)\,.
\end{equation}
We also suppose that $\gamma>\sqrt{\frac{2}{3M^{3}}}$. Equation
(\ref{brane-phi1}) defines the constant $C_{1}$
$$C_{1}=e^{\frac{\gamma\phi_{1}}{2}}\,.$$
The size of the extra dimension is
$$L=\frac{e^{\frac{\gamma\phi_{2}}{2}}-e^{\frac{\gamma\phi_{1}}{2}}}{k}\,.$$

Boundary conditions (\ref{b1}) result in
\begin{eqnarray}\label{F10}
F_{1}&=&\frac{4k}{\gamma}\,e^{-\frac{\gamma\phi_{1}}{2}}\,,\\
\label{F20}
F_{2}&=&-\frac{4k}{\gamma}\,e^{-\frac{\gamma\phi_{2}}{2}}\,.
\end{eqnarray}
We see that the form of (\ref{F10}), (\ref{F20}) is the same as
that of (\ref{F1}), (\ref{F2}). It follows from the fact that the
form of the solutions for the scalar field is the same for both
cases (see equations (\ref{ph1}) and (\ref{ph2})).

Boundary conditions (\ref{b3}) suggest that the brane tensions
$\epsilon_{1,2}$ should be also fine-tuned (compare with
(\ref{E1}), (\ref{E2}))
\begin{eqnarray}\label{E10}
\epsilon_{1}&=&-12M^{3}\frac{\beta\gamma}{\sqrt{6}}e^{-\frac{\gamma\phi_{1}}{2}}\,,\\
\label{E20}
\epsilon_{2}&=&12M^{3}\frac{\beta\gamma}{\sqrt{6}}e^{-\frac{\gamma\phi_{2}}{2}}\,.
\end{eqnarray}
\end{enumerate}
Thus, we have shown that for
$\sqrt{\frac{2}{3M^{3}}}<\gamma<2\sqrt{\frac{2}{3M^{3}}}$ there
exist two different solutions for the system with the bulk
potential (\ref{potential}). The only difference is in the
fine-tuned values of the brane tensions, see (\ref{E1}),
(\ref{E2}) and (\ref{E10}), (\ref{E20}). Very similar situation of
two different background solutions with a fixed bulk scalar field
potential was discussed in \cite{Cline:2000ky}. We note that the
background solutions presented above are exact.

A very peculiar case is $\gamma=\sqrt{\frac{2}{M^{3}}}$. For
$\gamma=\sqrt{\frac{2}{M^{3}}}$ we get fine-tuned tensions
\begin{eqnarray}\label{E100}
\epsilon_{1}&=&-4\beta\sqrt{3M^{3}}\,e^{-\frac{\phi_{1}}{\sqrt{2M^3}}}\,,\\
\label{E200}
\epsilon_{2}&=&4\beta\sqrt{3M^{3}}\,e^{-\frac{\phi_{2}}{\sqrt{2M^3}}}
\end{eqnarray}
for both cases ($H=0$ and $H\ne 0$)! It means that there are two
different stabilized solutions corresponding to the bulk scalar
field potential (\ref{potential}) and brane tensions (\ref{E100}),
(\ref{E200}). The first one is
\begin{equation}
A=\frac{1}{3}\ln\left(\sqrt{\frac{3}{M^{3}}}\beta
|y|+e^{\frac{\phi_{1}}{\sqrt{2M^{3}}}}\right)+C\,,\qquad
\phi=\sqrt{2M^{3}}\ln\left(\sqrt{\frac{3}{M^{3}}}\beta
|y|+e^{\frac{\phi_{1}}{\sqrt{2M^{3}}}}\right),
\end{equation}
\begin{equation}
H=0\,,\qquad
L=\sqrt{\frac{M^{3}}{3}}\left(\frac{e^{\frac{\phi_{2}}{\sqrt{2M^{3}}}}-e^{\frac{\phi_{1}}{\sqrt{2M^{3}}}}}{\beta}\right)\,,
\end{equation}
\begin{equation}
F_{1}=\sqrt{24}\beta e^{-\frac{\phi_{1}}{\sqrt{2M^{3}}}}\,,\qquad
F_{2}=-\sqrt{24}\beta e^{-\frac{\phi_{2}}{\sqrt{2M^{3}}}}\,;
\end{equation}
whereas the second solution is
\begin{equation}
A=\ln\left(\sqrt{\frac{1}{3M^{3}}}\beta
|y|+e^{\frac{\phi_{1}}{\sqrt{2M^{3}}}}\right)+C\,,\qquad
\phi=\sqrt{2M^{3}}\ln\left(\sqrt{\frac{1}{3M^{3}}}\beta
|y|+e^{\frac{\phi_{1}}{\sqrt{2M^{3}}}}\right),
\end{equation}
\begin{equation}
H=e^{C}\frac{\sqrt{2}\,\beta}{3\sqrt{M^{3}}}\,,\qquad
L=\sqrt{3M^{3}}\left(\frac{e^{\frac{\phi_{2}}{\sqrt{2M^{3}}}}-e^{\frac{\phi_{1}}{\sqrt{2M^{3}}}}}{\beta}\right),
\end{equation}
\begin{equation}
F_{1}=\sqrt{\frac{8}{3}}\beta
e^{-\frac{\phi_{1}}{\sqrt{2M^{3}}}}\,,\qquad
F_{2}=-\sqrt{\frac{8}{3}}\beta
e^{-\frac{\phi_{2}}{\sqrt{2M^{3}}}}\,.
\end{equation}
We see that there is a doubling of the background solution for
appropriate values of the model parameters. Both solutions
correspond to fixed sizes of the extra dimension,
maximally-symmetric spaces on the branes (maximally symmetric
spaces are Minkowski, $dS$ and $AdS$, see \cite{Weinberg}) and
both correspond to stationary cosmological solutions. Thus it
seems that there are no phenomenological criteria (such as a
symmetry criterion) to choose between the solutions.

Of course, such doubling is the consequence of our choice of the
stabilizing brane potentials (\ref{potent1}), (\ref{potent2}). One
can chose a more familiar form of the potentials (see, for
example, \cite{DeWolfe}):
\begin{eqnarray}\label{ftp1}
\lambda_{1}(\phi)=-4\beta\sqrt{3M^{3}}\,e^{-\frac{\phi_{1}}{\sqrt{2M^3}}}+Q_{1}\cdot
\left(\phi-\phi_{1}\right)+q_{1}^{2}\left(\phi-\phi_{1}\right)^{2},\\
\label{ftp2}\lambda_{2}(\phi)=4\beta\sqrt{3M^{3}}\,e^{-\frac{\phi_{2}}{\sqrt{2M^3}}}-Q_{2}\cdot
\left(\phi-\phi_{2}\right)+q_{2}^{2}\left(\phi-\phi_{2}\right)^{2},
\end{eqnarray}
where $Q_{1}$, $Q_{2}$, $q_{1}$, $q_{2}$ are constants. If
$Q_{1}=\sqrt{24}\,\beta e^{-\frac{\phi_{1}}{\sqrt{2M^{3}}}}$,
$Q_{2}=\sqrt{24}\,\beta e^{-\frac{\phi_{2}}{\sqrt{2M^{3}}}}$, then
one should consider the first background solution with $H=0$, if
$Q_{1}=\sqrt{\frac{8}{3}}\,\beta
e^{-\frac{\phi_{1}}{\sqrt{2M^{3}}}}$,
$Q_{2}=\sqrt{\frac{8}{3}}\,\beta
e^{-\frac{\phi_{2}}{\sqrt{2M^{3}}}}$, then one should consider the
second background solution with $H\ne 0$. But the latter brane
potentials appear to be more fine-tuned than potentials
(\ref{potent1}), (\ref{potent2}).

Now let us discuss the obtained results and compare them with
those obtained earlier. In \cite{DeWolfe} the problem of
uniqueness of solutions to (\ref{1})-(\ref{3}) was discussed.
Although the main solutions in \cite{DeWolfe} were obtained with
the help of the superpotential method (see also
\cite{Brandhuber:1999hb}), some statements for the general case
were also made. It was argued that there should exist a solution
to (\ref{1})-(\ref{3}) without fine-tuning of the parameters and
this solution is (locally) unique. This seems to be not correct in
general. Indeed, even with quite relaxed brane potentials
(\ref{potent1}), (\ref{potent2}), which contain additional degrees
of freedom described by the scalar fields $F_{1,2}(x)$, there
should be fine-tuning of the brane tensions (\ref{E1}), (\ref{E2})
or (\ref{E100}), (\ref{E200}) in order to get the solutions. For
the case $\gamma=\sqrt{\frac{2}{M^{3}}}$ the solutions appear to
be not unique globally, and only additional fine-tuning can
separate them (although the uniqueness of the solution was
discussed in \cite{DeWolfe} for the case $H\ne 0$, formally the
case $H=0$ should be included as a possible solution to this
system of equations). The subtle point is that our relaxed brane
potentials allow the existence of two different global solutions.
If we take highly fine-tuned brane potentials (\ref{ftp1}),
(\ref{ftp2}), which depend only on $\phi$ (analogous potentials
were considered in \cite{DeWolfe}), the solutions seem to be even
globally unique for the given values of $Q_{1}$, $Q_{2}$.

Nevertheless, it was noted in \cite{DeWolfe} that at least for
$H=0$ there can be a discrete set of solutions to the equations of
motion with fixed values of $\phi_{1}$ and $\phi_2$. As we have
seen above in a special case the solutions can even belong to
different classes -- one with $H=0$ and another with $H\ne 0$.

In this connection in should be mentioned that, as it was noted in
the beginning of the paper, in \cite{Cline:2000ky} an analogous
situation with two solutions corresponding to $H=0$ and $H\ne 0$
for a given bulk scalar field potential was considered. To find
the solutions the superpotential method was used, at the same time
the brane potentials were chosen to be stiff, which led to the
same physical consequences as our choice (\ref{potent1}),
(\ref{potent2}): the only boundary conditions for the scalar field
in both cases are (\ref{brane-phi1}), (\ref{brane-phi2}). In this
sense, besides the choice of the bulk scalar field potential, our
model and the model of \cite{Cline:2000ky} are very similar. The
first exact solution in \cite{Cline:2000ky} with $H=0$ was the one
previously obtained in \cite{DeWolfe}, whereas the second
approximate solution corresponding to $H\ne 0$ was found
perturbatively. It has been shown in \cite{Cline:2000ky} that the
second solution corresponds to the case where there is no
fine-tuning of the tension on the second brane (but with the
fine-tuning on the first brane retained). Our results show that
such situation can be realized even if we retain fine-tuning of
the second brane tension.

Finally we would like to note that although the doubling of
background solution happens only for a particular choice of the
parameter $\gamma$ in (\ref{potential}) and in the special case of
brane potentials (\ref{potent1}), (\ref{potent2}), more realistic
brane world models could also lead to different background
solutions for equal values of fundamental parameters. It is
necessary to bear this in mind while examining brane world models.

\section*{Acknowledgements}
The author is grateful to I.P.~Volobuev for valuable discussions.
The work was supported by grant of Russian Ministry of Education
and Science NS-1456.2008.2, FASI state contract 02.740.11.0244,
grant for young scientists MK-5602.2008.2 of the President of
Russian Federation, grant of the "Dynasty" Foundation, RFBR grant
08-02-92499-CNRSL-a and scholarship for young teachers and
scientists of M.V. Lomonosov

\end{document}